\documentclass[prl,twocolumn,showpacs]{revtex4}
\usepackage{epsfig}
\usepackage{color}

\newcommand{\be}{\begin{equation}}
\newcommand{\ee}{\end{equation}}
\newcommand{\bea}{\begin{eqnarray}}
\newcommand{\eea}{\end{eqnarray}}
\renewcommand{\r}{{\bf r}}

\newcommand{\br}{{\bf r}}
\newcommand{\brp}{{\bf r}^\prime}

\begin{document}


\title{ Spatial Log Periodic Oscillations of First-Passage Observables in Fractals}
\author{Eric Akkermans$^{1}$, Olivier Benichou$^{2}$, Gerald~V.~Dunne$^{3}$, Alexander Teplyaev$^{4}$ and Raphael Voituriez$^{2}$}

\affiliation{$^1$ Department of Physics, Technion Israel Institute of Technology, Haifa 32000, Israel
\\
$^2$ Laboratoire de Physique Th\'eorique de la Mati\`ere Condens\'ee (UMR 7600), Universit\'e Pierre et Marie Curie, 75255 Paris Cedex, France
\\
$^3$ Department of Physics, University of Connecticut, Storrs CT 06269, USA
\\
$^4$Department of Mathematics, University of Connecticut, Storrs CT 06269, USA}


\begin{abstract}
For transport processes in geometrically restricted domains, the mean first-passage time (MFPT) admits a general  scaling dependence on space parameters for diffusion, anomalous diffusion, and diffusion in disordered or fractal media. For transport in self-similar fractal structures, we obtain a new expression for the source-target distance dependence of the MFPT that exhibits both the leading power law behavior, depending on the Hausdorff and spectral dimension of the fractal, as well as small log periodic oscillations that are a clear and definitive signal of the underlying fractal structure. We also present refined numerical results for the Sierpinski gasket that confirm this oscillatory behavior.
\end{abstract}


\date{\today}

\maketitle



The  time taken for a random walker to reach for the first time a given target site from a given source site, is a particularly interesting physical observable that encodes information about a wide variety of physical and mathematical transport processes, ranging from chemical transport to propagation through networks \cite{redner}.
Recent progress has led to precise quantitative expressions that show how the mean first passage time (MFPT) is affected by the geometry [e.g., volume, shape, dimension] of the region in which the random walk occurs, even for fractals \cite{mfpt1,rvprl}. These results have been confirmed by  numerical simulations and have motivated various experimental studies \cite{dahan}. Here we argue that for such processes there are observable finer details encoded in the MFPT when the transport occurs on highly symmetric fractals: those possessing exact self-similar scaling, such as the Sierpinksi gasket. Technically, this result requires information concerning the off-diagonal Green's function $G({\bf r}_T |  {\bf r}_S)$ between the source and target points, and we show that for highly symmetric fractals, simple nearest neighbor  random walks generically lead to {\it spatial} log-periodic oscillations in first passage observables such as the MFPT. 

We recall briefly the essential results from \cite{mfpt1,rvprl}. Consider a random walker moving in a bounded region of $N$ sites.  In \cite{mfpt1,rvprl} it has been shown that the MFPT to a target site $\br_T$ starting from $\br_S $ satisfies in the large $N$ limit: 
\begin{eqnarray}
\langle {\bf T}\rangle/N\sim G(\br_T | \br_T)- G(\br_T | \br_S).
\label{approx}
\end{eqnarray}   
Note that the {\it r.h.s.} is independent of the confinement and involves the {\it  infinite space} Green's function $G({\bf r} | {\bf r}^\prime)$    defined by 
\begin{equation}
G({\bf r} | {\bf r}^\prime)=\int_0^\infty P({\bf r}, t | {\bf r}^\prime)\, dt,
\label{green}
\end{equation}
where the propagator $P({\bf r}, t | {\bf r}^\prime)$ gives the probability density that a random walker evolving in infinite space is  at site ${\bf r}$ at time $t$, starting from site ${\bf r}^\prime$ at time $0$. This probability density  satisfies the diffusion equation $\partial_t P({\bf r}, t | {\bf r}^\prime)=\Delta_\br P({\bf r}, t | {\bf r}^\prime)$.

We show in this paper that for certain highly symmetric fractal structures, the Green's function when averaged ($\overline{ \cdots} $) over source and target points, has the general scaling form:
\be
\overline{ G(\r_T | \r_S)} =  \frac{1}{r^{d_f - d_w}} \, {\cal G} \left( 2 \pi \frac{ \ln r}{\ln l} \right) \, .
\label{result}
\ee
where $r=|\br_T-\br_S |$ is the source-target distance, 
and $l$ is a decimation constant specific to the fractal (see Table 1 in \cite{ADT1} for some examples). The function ${\cal G} (x+1) = {\cal G} (x)$ is a periodic function of period unity. The fractal (Hausdorff) dimension $d_f$ and the walk dimension $d_w$ \cite{rammal,bass-carpet} are related to the spectral dimension as $d_s=2 d_f/d_w$. The walk dimension $d_w$ accounts for the anomalous space-time dispersion on fractals, $
 \langle r^2 (t) \rangle \sim t^{2/d_w}$, generalizing the usual expression, $\langle r^2 (t) \rangle \sim t$, for Euclidean manifolds \footnote{$\langle \cdots \rangle$ denotes the average over the probability distribution $P_t (\r, \r')$.}.
The scaling behavior of $\overline{G(\r_T |\r_S)}$ in (\ref{result}) has interesting consequences for a number of relevant physical quantities. Adapting the arguments of \cite{mfpt1,rvprl}, we deduce 
{\begin{eqnarray}
\langle {\bf T}\rangle/N\sim
\begin{cases}
{A-r^{d_w-d_f}{\cal G} \left( 2 \pi \frac{ \ln r}{\ln l} \right)\quad ,\quad d_w<d_f \cr
A+\ln r\, {\cal G} \left( 2 \pi \frac{ \ln r}{\ln l} \right)\qquad , \qquad d_w=d_f \cr
r^{d_w-d_f}{\cal G} \left( 2 \pi \frac{ \ln r}{\ln l} \right)\qquad ,\qquad d_w>d_f }
\end{cases}
\label{result1}
\end{eqnarray}
where $A$ is independent of both $N$ and $r$. Previous work \cite{mfpt1,rvprl} took ${\mathcal G}=c$, a constant,
and found excellent agreement with numerical results for a wide variety of diffusion processes: sub-diffusive ($d_w>d_f$), super-diffusive ($d_w< d_f$), and critical ($d_w=d_f$).

However, closer inspection of the numerical results in \cite{mfpt1,rvprl} indicates that for certain highly symmetric fractals there are small oscillatory deviations from the leading forms from (\ref{result1}) with ${\mathcal G}=c$. We argue here that these oscillatory deviations are not numerical artifacts, but are in fact real physical results that are strong indications of the underlying scaling and fractal structure of the diffusion process. We find an explicit approximate form for the function ${\mathcal G}$ which quantitatively explains these small oscillatory effects for the Sierpinski gasket.

It is clear from (\ref{approx}) that we need information about the spatially off-diagonal propagator. The approximation of taking ${\mathcal G}=c$ coincides with assuming a scaling form $P({\bf r}, t | {\bf r}^\prime) = \Pi\left(\frac{|\br-\brp|}{t^{1/d_w}}\right)/t^{d_f/d_w}$ \cite{havlin}. But we can improve on this estimate. Upper and lower bounds for the propagator $P(\br, t | \brp)$ have been proven by Barlow and Perkins \cite{bp} for a wide class of diffusion processes:
\be
{\cal F}_t (c_1 , c_2 , \r , \r') \leq P (\r, t | \r' ) \leq {\cal F}_t (c_3 , c_4 , \r , \r')
\label{p-bounds}
\ee
with 
\be
{\cal F}_t (c_i , c_j , \r , \r') \equiv {c_i \over  t^{d_s /2}}  \exp \left[ - c_j \left( {|\r - \r'|^{d_w} \over t} \right)^{{1 \over d_w - 1}} \right]
\label{scaling2}
\ee
for  real and positive constants $c_1$, \dots, $c_4$. The bounds (\ref{p-bounds}) translate into corresponding upper and lower bounds for the Green's function (here for $d_f \neq d_w$):
\footnote{Note: The notion of distance between any two points $\r$ and $\r'$ on a fractal, needs to (and can) be properly defined; here we use the more standard ``Euclidean'' notation $|\r - \r'|$ throughout the paper, in order to avoid the proliferation of notation. We can define the chemical distance for numerical purposes.}
\begin{eqnarray}
 \frac{c_5}{| \r - \r' |^{d_f - d_w}}
 \leq G( \r | \r' ) \leq  
 \frac{c_6}{| \r - \r' |^{d_f -d_w}}
\label{g-bounds}
\end{eqnarray}
Unfortunately, little is known rigorously about the constants $c_1, \cdots, c_6$, or about how $P (\r, t | \r' )$ or $G(\r | \r' )$ behave between the respective bounds (\ref{p-bounds}) and (\ref{g-bounds}).

In contrast to  these spatially {\it off-diagonal} quantities $P(\r, t | \r' )$ and $G(\r | \r' )$, much more is known about  the spatially {\it diagonal} propagator, $P(\br, t| \br)=\langle \br |e^{t\, \Delta}|\br\rangle$, (which when integrated over space gives the heat kernel trace):
\begin{eqnarray}
P(\br, t| \br)&=&  \frac{1}{t^{d_f/d_w}}\, F \left( { 2 \pi \ln t \over d_w \ln l } \right)
\label{diagonal}
\\
\hskip -1cm &\sim&
 \frac{c_7}{t^{d_f/d_w}}\left[1+c_8\, \cos\left( \frac{2\pi \ln t}{d_w \ln l}+\phi
\right)+\dots\right] \nonumber
\end{eqnarray}
The function $F(x) = F(x+1)$ is periodic  of period unity, which often is well approximated by its first harmonic \cite{ADT1}, and 
$l$ is the same decimation constant appearing in (\ref{result}). The log-periodic oscillations in the time variable $t$ are ubiquitous and already well recognized not only for diffusion on fractals, but also in other complex systems having a discrete (lacunar) scaling symmetry \cite{jullien}. 
The origin of these log-periodic oscillations in $t$ can be traced directly to spectral properties of the Laplacian $\Delta$. To see this, write $P(\br, t| \br)$ in terms of  the associated zeta function $\zeta_\Delta (s)$ defined by the inverse Mellin-Laplace transform \footnote{The Mellin-Laplace transform of the heat kernel is defined as $\zeta_\Delta (s) = {1 \over \Gamma (s)} \int_0^\infty dt \, t^{s-1} {\rm tr}_\br \left(e^{t\, \Delta}\right)$.}:
\begin{eqnarray}
{\rm tr}_\br \left(e^{t\, \Delta}\right)=\frac{1}{2\pi i}\int_C t^{-s}\, \Gamma(s)\, \zeta_{\Delta}(s)\, ds.
\label{zeta}
\end{eqnarray}
For symmetric fractals \cite{ADT1},  $\zeta_\Delta (s)$ has a tower of poles at $s_n=d_f/ d_w+2\pi i n/(d_w  \ln l)$ in the complex $s$ plane, which lead to  the log-periodic oscillatory form in (\ref{diagonal}).
Physically, these complex poles are a consequence of the {\it exponential} growth of the degeneracies and eigenvalues for the Laplacian on a fractal \cite{sasha-eg}. To have a better understanding of how this works, it is interesting to go back to the familiar case of a regular $d$-dimensional Euclidean manifold \cite{weyl-review,molchanov} where  the Laplacian eigenvalues have instead a {\it polynomial} growth, $\lambda_n\sim n^2$,  while the degeneracy factor scales like ${\rm deg}_n \sim n^{d-1}$, for a symmetric manifold, such as a $d$-dimensional hypersphere
\cite{minak}. 
Thus the zeta function goes like $\zeta_\Delta (s)\sim \sum_n n^{d-1}/n^{2s}\sim \zeta_R(2s-d+1)$, ($\zeta_R$ is the Riemann zeta function) which has the familiar real pole at $s=d/2$. This leads, via the Mellin transform, to the well known behavior of  the diagonal propagator, $P(\br, t| \br)\sim \sum_n n^{d-1}\, e^{-n^2\, t} \sim 1/t^{d/2}$, at short times.

On the other hand, for a symmetric fractal, with exponentially growing degeneracies and eigenvalues, 
\be
{\rm deg}_n\sim a^n\quad, \quad \lambda_n\sim b^n
\label{expon}
\ee
we find $\zeta(s)\sim \sum_n a^n/b^{n\,s}\sim 1/(1-a/b^s)$, which has a tower of complex poles with real part $s=\ln a/\ln b$, and vertical spacing $2\pi/\ln b$, which identifies $\ln b=d_w\, \ln l$, and $\ln a=d_f\, \ln l$, so that $\ln a/\ln b =d_f/d_w$. The complex poles produce the log-periodic oscillations in (\ref{diagonal}). 
For 
such a fractal,  
$P(\br, t | \br)\sim \sum_n a^n\, e^{-b^n\, t}\sim F\left(\frac{2\pi}{\ln b}\ln t\right) /t^{\ln a/\ln b}$, where $F$ is a periodic function of $\ln t$, as in (\ref{diagonal}).
For mathematical discussions of oscillations in heat kernel estimates see \cite{Hambly2010,Kajino2010}. In particular, there is a  class of fractals where oscillations are related not to the high degeneracies of eigenvalues but rather to large gaps in the spectrum. This topic is a subject of active current research \cite[and references therein]{St2005gaps,HSTZ}. 


Since space and time are coupled scaling variables related through the walk dimension $d_w$, we might expect to observe an analogous type of log-periodic oscillations for the stationary Green's function $G(\r | \r')$, as a function of the spatial distance between the points. 
To see this, consider an eigenfunction expansion:
\be
G(\r | \r' ) = \sum_n {\rm deg}_n {\phi_n ^* (\r') \phi_n (\r) \over \lambda_n}
\label{green2}
\ee
On a regular Euclidean manifold, when averaged ($\overline{\cdots }$) over the points $\r$ and $\r'$ we find $\overline{ \phi_n ^* (\r') \, \phi_n(\r )} \sim 
\tilde{f}(r/L_n)$, where $r$ denotes the distance $|\r -\r'|$, and $L_n$ depends on $n$. Consistency with scaling and  conservation of probability determine $L_n\sim 1/n$, using the polynomial growth of degeneracies and eigenvalues, as discussed above for the diagonal propagator. We recognize $1/L_n=\sqrt{\lambda_n}$ as the momentum, reflecting the usual quadratic dispersion relation, $k_n \sim \sqrt{\lambda_n}$. On a fully symmetric (or unbounded) manifold such as a $d$-dimensional hypersphere, this behavior holds without averaging. If $\tilde{F}(s)$ is the Mellin-Laplace transform of $\tilde{f}(r)$, then $\tilde{f}(r) = {1 \over 2i \pi} \int_C ds \, \tilde{F}(s) / r^s$, and 
\begin{eqnarray}
\overline{ G(\r, \r' )} &\sim& \sum_n n^{d-3} \frac{1}{2\pi i}\int_C ds\, \frac{\tilde{F}(s)}{(n \, r)^s} \nonumber \\
&=& \int_C { ds \over 2\pi i} \, \frac{\tilde{F}(s)}{r^s}\, \zeta_R(s-d+3)\sim\frac{1}{r^{d-2}}
\label{regularg}
\end{eqnarray}
which is the familiar form of the  Green's function, for $d\neq 2$, and is seen to result from the pole of the Riemann zeta function at $s=d-2$.


To generalize this result to fractals, we note the empirical numerical result that averaging over target and source points on fractals also leads to an averaged form  $\overline{ \phi_n ^* (\r) \, \phi_n(\r') }  \sim 
f( r/L_n)$, for some $L_n$. Consistency with scaling and the conservation of probability determines $L_n=b^{- n/d_w}= a^{- n/ d_f} = 1/l^n$, now using the exponential growth of degeneracies and eigenvalues, as discussed above for the diagonal propagator. 
Thus we find a scaling relation that involves the anomalous walk dimension $d_w$, and (for the non-critical case, $d_f\neq d_w$, i.e.  $d_s\neq 2$):
\begin{eqnarray}
\overline{G(\br_T | \br_S)}& \sim& \sum_n \frac{a^n}{b^{n}} \frac{1}{2\pi i}\int ds\, \frac{\tilde{F}(s)}{(b^{n/d_w} \, r)^s}\nonumber\\
&=&\frac{1}{2\pi i}\int ds\, \frac{\tilde{F}(s)}{r^s}\, \frac{1}{1-\frac{a}{b^{1+s/d_w}}}
\label{fractalg} 
\end{eqnarray}
which leads immediately to the scaling in (\ref{result}).

When comparing (\ref{result}) with the time-dependence of the diagonal propagator in  (\ref{diagonal}), we notice that the arguments of the periodic functions differ by a factor $1 / d_w$, which 
is consistent with the anomalous scaling of distance and time for diffusion on  a fractal.
Thus, we argue that the scaling in (\ref{result}), and the associated log periodic oscillations, have the same physical origin as their temporal counterpart in the diagonal propagator $P(\r, t | \r)$ in (\ref{diagonal}), coming from the exponential behavior  of the degeneracies ${\rm deg}_n$ and eigenvalues $\lambda_n$. 

The result (\ref{fractalg}) can be viewed as a refinement of the Green's function bounds in (\ref{g-bounds}), in a similar sense to the refinement (\ref{diagonal}) of the diagonal propagator bounds in the $t$ variable in (\ref{p-bounds}, \ref{scaling2}). The expression of $\overline{ G(\r | \r')}$ in (\ref{fractalg}) is consistent with (but more explicit than) the Barlow-Perkins bounds (\ref{p-bounds}). It is also consistent with rigorous results in \cite{kumagai,benarous,grigoryan} for the small $t$ behavior of $t^{d_f/d_w}\ln P (\r, t| \r')$, which behaves also as some log-periodic function of $r$. 


The expression (\ref{result}) for the averaged stationary Green's function constitutes the main result of this Letter. We now present a numerical test of its consequences for the mean first passage time (MFPT) $\langle {\bf T}\rangle$. A number of recent works \cite{mfpt1,rvprl} have derived precise expressions for the MFPT in the limit of large domains.  
Motivated by the case of the diagonal propagator (or heat kernel) studied in \cite{ADT1}, we propose to approximate the log periodic function ${\mathcal G}$ in (\ref{result}) by its first harmonic:
\be
{\mathcal G}\sim \left[1+b_2\cos\left(\frac{2\pi\, \ln r}{\ln l}+\phi\right)+\dots \right] 
\label{harmonic}
\ee
Inserting this expression for ${\mathcal G}$ into (\ref{result1}) we obtain new expressions for the MFPT $\langle {\bf T}\rangle$, which generalize those found in \cite{mfpt1,rvprl}. 
\begin{figure}[htb]
\includegraphics[scale=0.25]{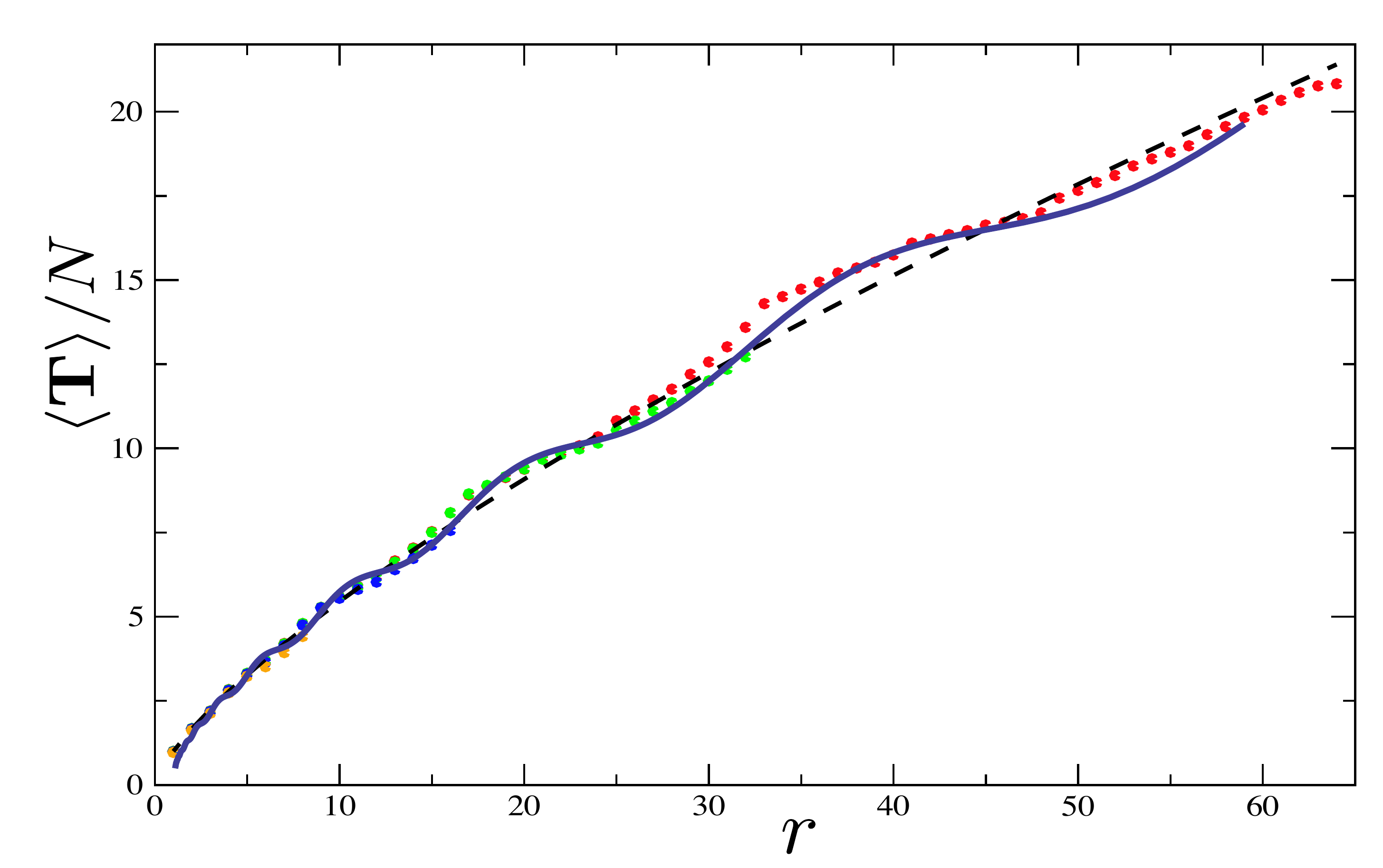}
\caption{Plot of the spatial dependence of the MFPT observable $\langle {\bf T} \rangle/N$, as a function of the source-target distance $r$, for the Sierpinski gasket. The dotted points are numerical data (gaskets of generation 3,4,5,6 respectively in yellow, blue, green, red),  the dashed black line is the leading behavior $r^{d_w-d_h}$, as studied in \cite{mfpt1,rvprl}. The  blue solid line  is a more refined fit from (\ref{harmonic}) with $b_2=0.05$, and $\phi=\pi/2$.}
\label{fig1}
\end{figure}
\begin{figure}[htb]
\includegraphics[scale=0.25]{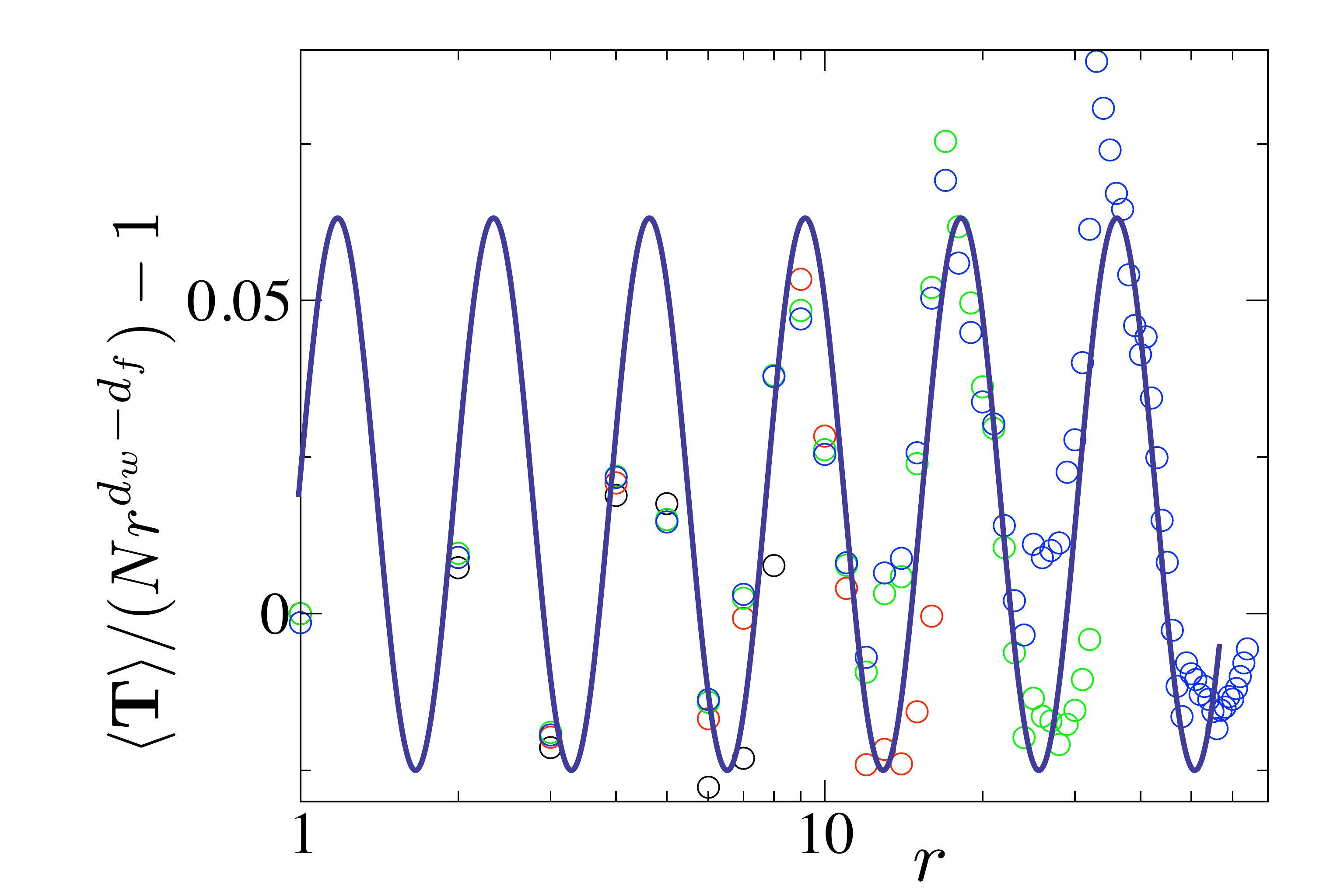}
\caption{Plot of the corrections to the leading $r^{d_w-d_h}$ behavior of the spatial dependence of the MFPT $\langle {\bf T}\rangle/N$, as a function of the source-target distance $r$, for the Sierpinski gasket. Circles stand for numerical simulations (gaskets of generation 3,4,5,6 respectively in black, red, green, blue), and the  blue solid line  is a  fit from (\ref{harmonic}). This figure clearly shows the log-periodic oscillations with the variable $2\pi \ln r/\ln l$.}
\label{fig2}
\end{figure}
In Figures \ref{fig1} and \ref{fig2}, we compare this new expression for the MFPT with numerical data, for the Sierpinksi gasket, which has $d_f>d_w$. First, in Figure \ref{fig1} we use the numerical results of \cite{mfpt1,rvprl}, and compare with the leading power-law behavior with ${\mathcal G}=1$, as well as with our refined log periodic oscillation form in (\ref{harmonic}).  We have used a simple fit with a single periodic function, with period $\ln l$ in the variable $\ln r$. The agreement with this refined form is very good. In Figure \ref{fig2}, we present the result of a new and more detailed numerical analysis for the corrections to the leading behavior, plotted as a function of the logarithmic separation distance $\ln r$. This plot exhibits clear log-periodic oscillations of corresponding period $\ln l=\ln 2$ for the Sierpinski gasket. 

As a further test of our analysis, one can note the absence of oscillations in the case of fractal trees (e.g. the T graph studied in    \cite{rvprl}). This is related to the fact, proven in the mathematics literature, that on trees the Green's function is essentially linear with respect to the effective resistance distance, in which case no spatial log-periodic oscillations are expected. The Green's function and effective resistance on trees were studied in \cite{kigami-dend}, and on more general fractals in  \cite{kigami}. The absence of oscillations was also checked numerically in the case of disordered fractals on the example of critical percolation clusters (see Figure 3 of \cite{rvprl}), which strongly suggests that their existence requires an exact decimation symmetry, found in deterministic fractals such as the the Sierpinski gasket. 


In conclusion, we have argued that a refinement of previously obtained bounds for the stationary Green's function $\overline{ G(\r | \r')}$ associated to the Laplacian on symmetric fractals can be found that is given by expression (\ref{fractalg}).  In addition to a  leading power law dependence already found in \cite{mfpt1,rvprl}, we have shown that $\overline{ G(\r | \r')}$ exhibits also small  log-periodic oscillations in the spatial variable. Our argument traces the physical origin of this refined functional form to the exponential growth of degeneracies and eigenvalues for the Laplacian on a  symmetric fractal, and is consistent with the anomalous dispersion for diffusion on a fractal. The resulting spatial form parallels the temporal behavior of the diagonal heat kernel, which is well-studied both physically and mathematically. While much less is known rigorously for the spatial dependence, our conjectured form has an immediate consequence for the mean first-passage time $\langle {\bf T}\rangle$, which is accessible to numerical and even experimental investigation. We have shown that our refined expression provides a good quantitative fit which also demonstrates unambiguously the existence of log-periodic oscillations. It would be very interesting to investigate the existence of similar behaviors for diffusion on other classes of complex scale-free networks, beyond the leading order results in \cite{mfpt1,rvprl}.


\medskip

This work was supported by US DOE under grant DE-FG02-92ER40716, by the Israel Science Foundation Grant No.924/09 and NSF. We thank J. Kigami and N. Kajino for helpful discussions and correspondence and E. Gurevich for relevant remarks.

    \end{document}